
\input harvmac
\noblackbox

\Title{UCLA/92/TEP/15}
{\vbox{\centerline{Operator Coefficients for}
\vskip2pt\centerline{Composite
Operators in the $(\phi^4)_4$ Theory$^*$}}}
\footnote{}{*This work was supported in part by the U.S.
Department
of Energy, under Contract DE-AT03-88ER 40384 Mod A006
Task C.}

\centerline{Hidenori SONODA}
\bigskip\centerline{\it Department of Physics,
UCLA, Los Angeles, CA 90024-1547, USA}

\vskip 1in
In a previous paper we derived a relation
between the operator product coefficients
and anomalous dimensions.
We explore this relation in the $(\phi^4)_4$ theory
and compute the coefficient functions in the
products of $\phi^2$ and $\phi^4$ to first order
in the parameter $\lambda$.   The calculation
results in two-loop beta functions.

\Date{May 92}

\def\phifour{{\phi^4 \over 4!}}
\def\phitwo{{\phi^2 \over 2}}
\def\O{{\cal O}}
\def\Ox{\O_{\lambda}}
\def\Om{\O_m}
\def\R{{\cal R}}
\def\one{{\bf 1}}
\def\g1{g_\one}
\def\b1{\beta_\one}
\def\bm{\beta_m}
\def\bx{\beta_\lambda}
\def\dl{{d \over dl}}
\def\vevcx#1{\langle #1 \rangle^c_{m^2,\lambda}}
\def\vevx#1{\langle #1 \rangle_{m^2,\lambda}}
\def\vevc#1{\langle #1 \rangle^c_{m^2,0}}
\def\vev#1{\langle #1 \rangle_{m^2,0}}
\def\ep{\epsilon}
\def\On{\O_{i_1} (r_1) ... \O_{i_n} (r_n)}
\def\x{\lambda}
\def\e{{\rm e}}
\def\D{\Delta}

\newsec{Introduction}

One popular view of renormalization is to regard
divergences that we encounter in perturbative calculations
as a consequence of the use of unphysical parameters.
Once we use physical parameters, all physical quantities
we calculate are rendered finite in terms of them.
There is nothing wrong with this view, but it is not
satisfactory.  The perturbative divergences should not
occur for no reason, and we would
like to understand why they appear.

One clue is provided by the dimensional regularization with
the minimal subtraction.  In the $\phi^4$ theory
defined in $D = 4 - \ep$-dimensional euclidean space
the bare parameter $\x_0$ is related
to the renormalized parameter $\x$ by
\eqn\erenorm{\x_0 = Z(\x;\ep) \x \mu^\ep ,}
where $\mu$ is a renormalization point.
We can expand the renormalization constant $Z(\x;\ep)$ as
\eqn\eexpansion{Z(\x;\ep) = 1 + {1 \over \ep}~Z^{(1)} (\x)
+ {1 \over \ep^2}~Z^{(2)} (\x) + ...~.}
The simple pole is related directly to the beta function
\eqn\erelation{\bx (\x) = - \x^2 {d \over d\x}~ Z^{(1)} (\x) ,}
and the higher order poles are

related to the simple pole.\ref\rthooft{G. 't Hooft, Nucl.
Phys. B62(1973)444}  Thus, a nonvanishing beta function
implies divergences to be subtracted in perturbative
calculations.

In a previous paper \ref\rqcd{H.~Sonoda,

``Composite operators in QCD'', UCLA/91/TEP/47}
we have introduced a variational formula
that realizes the derivative of
correlation functions with respect to
a parameter ($\x$ in the case of the $(\phi^4)_4$
theory) in terms of an insertion of a conjugate operator.
The variational formula is formulated in the coordinate
space, and it involves divergent subtractions,
which are due to short-distance singularities
of the {\it renormalized} theory.  We can

use the variational formula recursively
to evaluate higher order derivatives.  By evaluating
the higher order derivatives at $\x = 0$,
we reproduce perturbation theory.  Hence, with
the use of the variational formula we can relate
the divergences that we must subtract in perturbative
calculations to the short-distance singularities
of the renormalized theory.  This is a more
physical analogue
of what we observed in the dimensional regularization

with the minimal subtraction.
Our way of doing perturbative calculations
is more in the spirit of

ref.~\ref\rwilson{K. Wilson, Phys. Rev. 179(1969)1499}
than the standard diagrammatic way.

In ref.~\rqcd\ we were mainly concerned with arbitrary
composite operators in QCD.  Here we will concentrate on
the scalar operators conjugate to renormalized
parameters in the context of perturbative
$\phi^4$ theory in four-dimensional euclidean space.
In particular we will give a physical interpretation
of the beta functions as operator product coefficient
functions,
and we will apply the variational formula of ref.~\rqcd\
for the first order perturbation theory.

This paper is organized as follows.  In sect.~2 we will
summarize the relevant results of ref.~\rqcd.  Then, in
sect.~3, we introduce a particular choice of parameters,

for which the beta functions determine uniquely the operator
product coefficient functions of the conjugate operators.
We apply the variational formula of ref.~\rqcd\ to the
operator product coefficients in sect.~4.  This result
will be used in sect.~8 to calculate the first order
corrections to the coefficient functions.  These corrections
provide with, among others, beta functions up to second
order in $\x$, using the relation derived in sect.~3.

Sects.~5, 6, and 7 are preparations for the calculations
in sect.~8.  Finally in sect.~9 we conclude the paper.

\newsec{Review of previous results}

In this section we review the relevant results from a previous
paper \rqcd, which was written for QCD in

four-dimensional euclidean space.  We will give a summary
with appropriate modifications for the
$(\phi^4)_4$ theory.

The $(\phi^4)_4$ theory has three parameters
$\g1$, $m^2$, and $\x$ conjugate to the operators
$\one$, $\Om$ (corresponding to
$\phitwo$), and $\Ox$ (corresponding to
$\phifour$), respectively.

The renormalization group (RG) is parameterized
by a real number $l$, and the RG transformation
corresponding to $l$ maps the distance $r$ to $r \e^{-l}$,
while the renormalization point is always fixed at $r=1$.
The $l$-dependence of the parameters is given by
the RG equations
\eqn\ebeta{\eqalign{\dl~\g1 &= 4 \g1 +

(m^2)^2 \b1 (\x )\cr
\dl~ m^2 &= (2 + \bm (\x ))~ m^2\cr
\dl ~\x &= \bx (\x ) ,\cr}}
where the beta functions can be expanded as
\eqn\ebetaTaylor{\eqalign{\b1 (\x) &= b_{\one,0} + b_{\one,1}
\x + ... \cr\bm (\x) &= b_{m,1} \x + {b_{m,2} \over 2} \x^2 + ...
\cr\bx (\x) &= {b_{\x,1} \over 2} \x^2 + {b_{\x,2} \over 3!} \x^3
+ ... ~.\cr}}
The parameter $l$ grows towards infrared, and our
beta functions differ from the standard ones by sign.

Let $\g1 (l;\g1,m,\x)$, $m^2(l;m,\x)$, and $\x(l;\x)$ be the

running parameters that satisfy
the initial conditions
\eqn\einitial{\g1(0;\g1,m^2,\x) = \g1 ,~ m^2(0;m^2,\x) = m^2, ~
\x(0;\x) = \x .}
Note that $\g1 (\ln r), m^2 (\ln r)$, and $\x(\ln r)$ are invariant
under
the RG.  (We will suppress the dependence of the running
parameters
on $\g1, m^2$, and $\x$.)   To solve $\g1(l)$ and $m^2(l)$ in
terms
of $\x (l)$ we introduce
\eqn\eE{E(\x', \x) \equiv \exp \left[ \int_\x^{\x'} dx~{\bm(x)
\over
\bx(x)} \right] ,}
\eqn\eF{F(\x', \x) \equiv \int_\x^{\x'} dx~{\b1(x) \over \bx(x)}~
E(x,\x)^2 .}
Then we find
\eqn\eRGgonem{\eqalign{\g1 (l) &= \g1 + m^4 F(\x(l),\x) \cr
m^2 (l) &= m^2 E(\x(l),\x) .\cr}}

In ref.~\rqcd\ we considered the
RG and short-distance properties of

arbitrary scalar composite operators.
Under the RG an operator of scaling
dimension $x$ can mix only with operators
of scaling dimension $y$ not larger
than $x$, i.e., $y \le x$.  In this
paper we will be interested in the
scalar composite operators of dimension
not greater than four which are invariant
under the ${\bf Z}_2$ transformation
\eqn\eztwo{ \phi \to - \phi .}
In addition to the conjugate operators
$\one, \Om$, and $\Ox$ we have only
\eqn\eR{\R \equiv \partial^2 \Om}
as such operators.  The other possible
candidate is an operator
$\partial_\mu \phi \partial_\mu \phi$,
but it is equal to a linear combination of
$\Om$ and $\Ox$ by the equation of motion.

We now introduce a matrix notation in which
$\O$ stands for a four-dimensional column vector
$(\one, \Om, \Ox, \R)^T$.  Then, from \ebeta,
we can derive the RG for $\O$ in the matrix notation
as follows:
\eqn\eRG{ \dl \O (r) = (X + \Gamma (m,\x)) \O (r) ,}
where
\eqn\eXGamma{X \equiv {\rm diag} (0,2,4,4) ,~
\Gamma (m,\x) \equiv \left( \matrix{0&0&0&0\cr
- 2 m^2 \b1& - \bm&0&0\cr
- m^4 \b1'& - m^2 \bm'& - \bx'& 0 \cr
0&0&0&- \bm\cr} \right) .}
The RG equation for $\R$ can be obtained from
that for $\Om$.  We have assumed that $\Ox$ does
not mix with $\R$ under the RG.\foot{From the result
of \ref\rmix{H. Sonoda, Nucl. Phys. B366 (1991)
629}, this assumption is valid
if $2 b_{m,1}/b_{\x,1}$ is not a positive integer greater than
1.  We will see that  $2 b_{m,1}/b_{\x,1} = 1/3$; the condition
is satisfied.}

For later convenience we introduce a matrix $G(r;m,\x)$ that
satisfies
\eqn\eGeq{\dl G(r;m,\x) = (X + \Gamma (m,\x)) G(r;m,\x)}
and
\eqn\eGinitial{G(1;m,\x) = 1 .}
Using the explicit expressions of $X$ and $\Gamma$ given
by
\eXGamma, we find
\eqn\eG{G(r;m^2,\x) \equiv
\left( \matrix{1&0&0&0\cr
2 m^2 F(\x(\ln r),\x)&{1\over r^2} E(\x(\ln r),\x)&0&0\cr
m^4 \partial_\x F&{m^2 \over r^2} \partial_\x E&{\bx(\x(\ln r))
\over r^4
\bx(\x)}&0\cr
0&0&0&{1 \over r^4} E\cr} \right) ,}
where $E, F$ are defined by \eE, \eF.

We now define the singular part of the

coefficient functions

$C_\x , C_m$ (four by four matrices)

in operator product expansions (OPE) by
\eqn\eope{\eqalign{
\Om (r) \O (0) &= C_m (r; m^2, \x) \O (0)

+ o \left( {1 \over r^4} \right) \cr
\Ox (r) \O (0) &= C_\x (r; m^2, \x) \O (0)

+ o \left( {1 \over r^4} \right) , \cr}}
where the angular average over
the orientation of the coordinate vector $r_\mu$ is taken
so that only scalar operators can appear on the right-hand
sides.
Eqs.~\eRG\ imply
\eqn\eRGope{\eqalign{
\dl C_m (r;m^2,\x) &= (2-\bm(\x)) C_m + [X + \Gamma
(m^2,\x),
C_m] \cr
\dl C_\x (r;m^2,\x) &= (4-\bx'(\x)) C_\x - m^2 \bm'(\x) C_m +
[X + \Gamma (m^2, \x), C_\x] ,}}
which are be solved by
\eqn\eCmCx{\eqalign{
C_m (r;m^2,\x) &= {1 \over r^2}~E(\x(\ln r),\x) G(r;m^2,\x)

\cdot H_m (m^2(\ln r),
\x (\ln r)) \cdot G^{-1} ,\cr
C_\x (r;m^2,\x) &= {1 \over r^4} {\bx (\x(\ln r)) \over \bx (\x)} G
\cdot H_\x (m^2(\ln r), \x(\ln r)) \cdot G^{-1} \cr

&~~~~~~ + m^2 ~{\bm (\x(\ln r)) - \bm (\x) \over \bx(\x)}~C_m
(r;m^2,\x) .}}
Here we define
\eqn\eHdef{H_m (m^2,\x) \equiv C_m (1;m^2,\x) ,~
H_\x (m^2,\x) \equiv C_\x (1;m^2,\x) .}

In ref.~\rqcd\ we have introduced the following variational
formulas:
\eqn\evarm{\eqalign{
&- \partial_{m^2} \vevcx{\On} \cr
&~~~~~ = \lim_{\ep \to 0} \Big[ \int_{|r-r_i| \ge \ep} d^4 r~
\vevcx{\Om (r) \On} \cr
&~~~~~~~~ + \sum_{k=1}^n A_{mi_k,j} (\ep; m^2,\x)
\vevcx{\O_{i_1} (r_1) ... \O_j (r_k) ... \O_{i_n} (r_n)} \Big] \cr }}
\eqn\evarx{\eqalign{
&- \partial_\x \vevcx{\On} \cr
&~~~~~ = \lim_{\ep \to 0} \Big[ \int_{|r-r_i| \ge \ep} d^4 r~
\vevcx{\Ox (r) \On} \cr
&~~~~~~~~ + \sum_{k=1}^n A_{\x i_k,j} (\ep; m^2,\x)
\vevcx{\O_{i_1} (r_1) ... \O_j (r_k) ... \O_{i_n} (r_n)} \Big] ,\cr }}
where $^c$ denotes a connected part, and the subtractions

$A_\x, A_m$ (four by four matrices)
are defined as the sum of an integral of
a singular coefficient function and a finite counterterm:
\eqn\eAm{A_m (\ep; m^2,\x) \equiv - \int_{1 \ge r \ge \ep}
d^4 r~C_m (r;m^2,\x) + c_m (m^2,\x) }
\eqn\eAx{A_{\x} (\ep; m^2,\x) \equiv - \int_{1 \ge r \ge \ep}
d^4 r~C_{\x} (r;m^2,\x) + c_{\x} (m^2,\x) .}
We note that the variational formulas are valid

not only for $\O_i = \one, \Om, \Ox, \R$, but also for
any arbitrary operators.  The mass insertion in \evarm\
is familiar from Callan-Symanzik equations \ref\rcseqs{C. G.
Callan,
Phys. Rev. D2 (1970) 1541;

K. Symanzik, Comm. Math. Phys. 18(1970)227}.

We will explain the relation between the
mass insertion in the variational formula \evarm\ and that
in Callan-Symanzik equations in Appendix A.

The finite counterterms have the following structure:
\eqn\ecmcx{\eqalign{c_m (m^2,\x) &= \left(
\matrix{0&0&0&0\cr
c_{mm,\one} (\x)&0&0&0\cr
c_{m\x,\one} (m^2,\x)&c_{m\x,m} (\x)&0&0\cr
c_{m\R,\one} (m^2,\x)&c_{m\R,m} (\x)&0&0\cr} \right) \cr
c_\x (m^2,\x) &= \left( \matrix{0&0&0&0\cr
c_{\x m,\one} (m^2,\x)&c_{\x m,m} (\x)&0&0\cr
c_{\x\x,\one} (m^2,\x)&c_{\x\x,m} (m^2, \x)&c_{\x\x,\x} (\x)
&c_{\x\x,\R} (\x) \cr
c_{\x\R,\one} (m^2,\x)&c_{\x\R,m} (m^2,\x)&c_{\x\R,\x} (\x)&
c_{\x\R,\R} (\x) \cr} \right) . \cr }}
The finite counterterms are finite polynomials in
$m^2$; $c_{m\x,\one}, c_{\x m,\one}, c_{\x\x,m}, c_{\x\R,m}$
are first order, and $c_{\x\x,\one}, c_{\x\R,\one}$

are second order in $m^2$.

In each finite counterterm we call the part proportional

to the highest power of $m^2$ the {\it maximal} part.
As a consequence of the equality
\eqn\eeqaul{\partial_{m^2} \vevcx{\Ox} = \partial_\x
\vevcx{\Om},}
we obtain the symmetry
\eqn\esym{c_{m \x,i} = c_{\x m,i} .}

By imposing consistency between the variational formulas
\evarm, \evarx\ and the RG eqs.~\ebeta, \eRG, we obtain
the main result of ref.~\rqcd:
\eqn\eH{\eqalign{2 \pi^2 H_m (m^2,\x)

&= \partial_{m^2} \Phi (m^2,\x)
+ [c_m, \Phi] + \bx (\x) (\partial_\x c_m - \partial_{m^2} c_\x
+ [c_\x, c_m]) \cr
2 \pi^2 H_\x (m^2,\x) &= \partial_{\x} \Phi (m^2,\x)
+ [c_\x, \Phi] - m^2 (2+\bm(\x))

(\partial_\x c_m - \partial_{m^2} c_\x
+ [c_\x, c_m]) ,\cr}}
where
\eqn\ePhi{\Phi (m^2,\x) \equiv X + \Gamma (m^2,\x)

+ (2+\bm (\x)) m^2 c_m + \bx c_\x .}
Here $c_m, c_\x$ are the finite counterterms in
\ecmcx.

We can use the variational formulas \evarm, \evarx\
recursively
to evaluate higher order derivatives.  By imposing
commutativity
of the derivatives, we obtain the following constraint on
the finite counterterms:
\eqn\ecurv{\big(\partial_\x c_m - \partial_{m^2} c_\x
+ [c_\x, c_m]\big) \vevx{\O} = \Omega_{\x m} \vevx{\O} ,}
where the curvature $\Omega_{\x m}(m^2,\x)$ is defined by
\eqn\eOmega{\eqalign{& \Omega_{\x m} (m^2,\x) \vevx{\O}
\equiv
\int_{1 \ge r} d^4 r~ {\rm F.P.}~
\int_{1 \ge r'} d^4 r'~\cr
&~~~~~~~~~~ \vevcx{ \Om (r) (\Ox (r') \O(0) - C_\x (r') \O(0))

- \Ox (r) (\Om (r') \O(0) - C_m (r') \O(0)) }, \cr}}
where F.P. stands for the operation of subtracting
the unintegrable part under integration over $r$.

Eqs.~\eH\ imply that the OPE coefficients are
unambiguously
determined by the beta functions $\b1, \bm, \bx$
and finite counterterms $c_m, c_\x$.  But the converse is
not true.  While OPE describes the singularities
of only {\it two} operators close together, the finite
counterterms can describe the behavior of
{\it three} operators close together, as can be seen
from the definition of the curvature \eOmega.
We will see from the calculations in the following
sections that both \eH\ and \ecurv\ are necessary for
the determination of the beta functions and finite
counterterms.

Finally by substituting \eH\ into \eCmCx,

we can find the subtractions
$A_m$ \evarm\ and $A_\x$ \evarx\ in

terms of the finite counterterms:
\eqn\eA{\eqalign{
A_m (\ep;m^2,\x) &= - G(\ep;m^2,\x) \left( \partial_{m^2} +
\ep^2 E(\x(\ln \ep),\x) ~c_m (m^2(\ln \ep), \x(\ln \ep)) \right)
G^{-1} \cr
A_{\x} (\ep;m^2,\x) &= - G(\ep;m^2,\x) \Big( \partial_\x

+ m^2 \ep^2 \partial_\x E(\x(\ln \ep),\x) ~c_m

(m^2(\ln \ep),\x(\ln \ep)) \cr

&~~~~~ +
{\bx(\x(\ln \ep)) \over \bx(\x)} ~c_\x (m^2(\ln \ep),\x(\ln \ep))
\Big)
G^{-1} .\cr}}
{}From this formula one can see that
non-maximal counterterms give only divergent subtractions,
while maximal counterterms give both divergent and finite
subtractions.\rqcd\

\newsec{Choice of parameters}

The operators $\one$, $\Om$, and $\Ox$ are conjugate to
$\g1$, $m^2$ and $\x$, and the precise definitions of

$\Om, \Ox$ depend
on the choice of the parameters.  As has been shown
in ref.~\rqcd\ (sect.~7), under

the most general reparametrization
\eqn\erepara{\eqalign{
\tilde{\g1} &= \g1 + {1 \over 2}~m^4 A(\x) \cr
\tilde{m^2} &= m^2 B(\x) \cr
\tilde{\x} &= f(\x) \cr}}
the conjugate operators transform as
\eqn\etrans{(\one, \tilde{\Om}, \tilde{\Ox})^T
= N(m^2,\x)  ~(\one, \Om, \Ox)^T ,}
where
\eqn\eN{N(m^2,\x) = \left( \matrix{
1&0&0\cr
- m^2 {A \over B}&{1 \over B}&0\cr
m^4 {A \over f'} \left({B'\over B} - {A' \over 2 A}\right)&

- {m^2 \over f'} {B' \over B}&{1 \over f'}\cr} \right) .}

The corresponding changes in the finite counterterms are
given by
\eqn\ecchange{\eqalign{
\tilde{c_m} &= {1 \over B} N \cdot (c_m + \partial_{m^2})

N^{-1} \cr
\tilde{c_\x} &= {1 \over f'} N \cdot (c_\x + \partial_\x)

N^{-1}

- m^2 {B' \over f' B}~ N \cdot (c_m + \partial_{m^2}) N^{-1}
.\cr}}
We find especially that
\eqn\especial{\eqalign{(\tilde{c_m})_{m,\one} &= {1 \over
B^2}~
(c_{mm,\one} + A) \cr

(\tilde{c_m})_{\x,m} &= {1 \over f'}~
(c_{m\x,m} + {B' \over B} ) \cr
(\tilde{c_\x})_{\x,\x} &= {1 \over f'}~
(c_{\x\x,\x} + {f' \over f}) .\cr}}

Hence, by choosing
\eqn\echoice{\eqalign{A &= - c_{mm,\one} \cr
B &= \exp \left[ - \int_0^\x dx~c_{m\x,m} (x) \right] \cr
f' & = \exp \left[ - \int_0^\x dx~ c_{\x\x,\x} (x) \right]~~
(f(0) = 0) ,\cr}}
we can satisfy the conditions
\eqn\egauge{\tilde{c_m}_{m,\one} = 0,~
\tilde{c_m}_{\x,m} = 0,~ \tilde{c_\x}_{\x,\x} = 0 .}
In other words the above conditions \egauge\ specify the
choice
of the parameters $\g1, m^2$, and $\x$ uniquely.  In the
following
we will use these special parameters and omit the tildes.

Under the gauge condition \egauge\ three coefficient
functions
become extremely simple.  From \eCmCx\ and \eH\ we find
\eqn\ethree{\eqalign{C_{mm,\one} (r;\x) &= - {1 \over \pi^2
r^4}~
\b1 (\x(\ln r)) E(\x(\ln r),\x)^2 \cr
C_{m\x,m} (r;\x) &= - { 1 \over 2 \pi^2 r^4}~
{\bx (\x(\ln r)) \over \bx (\x)}~\bm' (\x(\ln r)) \cr
C_{\x\x,\x} (r;\x) &= - {1 \over 2 \pi^2 r^4}

{\bx (\x(\ln r)) \over \bx (\x)}~\bx'' (\x(\ln r)) .\cr}}
These three coefficient functions are completely determined
by the three beta functions $\b1, \bm$, and $\bx$.
Our choice of the parameters are special in the sense that
the corresponding beta functions are related to the physical
singularities of the conjugate operators at short distances.
\foot{Since the $(\phi^4)_4$ theory is not well-defined
at short distances, this statement is valid only within
perturbation theory.}

\newsec{Variational formula for OPE coefficients}

As an application of the variational formulas \evarm, \evarx\
we will derive a variational formula for OPE
(operator product expansion) coefficients
in this section.

We consider a general OPE
\eqn\eopegeneral{\O_i (r) \O_j (0) = C_{ij,k} (r;m^2,\x) \O_k
(0) ,}
where $\O_i$'s are not necessarily the operators
$\one, \Om, \Ox, \R$.  (The summation over $k$
is implied.)  The variational formula \evarx\ gives
\eqn\evarope{\eqalign{- \partial_\x \langle \O_i (r) \O_j (0)
\rangle_{m^2,\x}
& = \lim_{\ep \to 0} \Big[ \int_{\scriptstyle |r' - r| \ge \ep
\atop\scriptstyle r'  \ge \ep} d^4 r'~ \langle (\Ox(r') - \langle
\Ox \rangle)
\O_i (r) \O_j (0) \rangle_{m^2,\x} \cr
& ~~~ + A_{\x i ,k} (\ep) \langle \O_k (r) \O_j (0)
\rangle_{m^2,\x}

+ A_{\x j,k} (\ep) \langle \O_i (r) \O_k (0) \rangle_{m^2,\x} \Big]
.\cr}}
By applying the variational formula to $\langle \O_k
\rangle_{m^2,\x}$
and using the OPE \eopegeneral\ we find
\eqn\evarope{\eqalign{
& ~~~ - \partial_\x C_{ij,k} (r;m^2,\x) \cdot \vevx{\O_k} \cr
&= \lim_{\ep \to 0} \Big[ \int_{\scriptstyle |r' - r| \ge \ep
\atop\scriptstyle r'  \ge \ep} d^4 r'~ \langle (\Ox(r') - \langle
\Ox \rangle)
(\O_i (r) \O_j (0) - C_{ij,k} (r) \O_k (0)) \rangle_{m^2,\x} \cr
&~~~ + ( A_{\x i,k} (\ep) C_{kj,l} (r) + A_{\x j,k} (\ep) C_{ik,l} (r)

- C_{ij,k} (r) A_{\x k,l} (\ep) ) \langle \O_l \rangle_{m^2,\x} \Big]
.\cr}}

We note that for any $r_0 > r$
\eqn\ezero{\int_{r' \ge r_0} d^4 r' ~
\langle (\Ox (r) - \langle \Ox \rangle_{m^2,\x}) (\O_i (r) \O_j (0)
- C_{ij,k} (r) \O_k (0)) \rangle_{m^2,\x} = 0 }
as a consequence of the OPE \eopegeneral.  Hence, for an
arbitrary
$r_0$ and arbitrary $r < r_0$ we find
\eqn\evaropetwo{\eqalign{
& ~~~ - \partial_\x C_{ij,k} (r;m^2,\x) \cdot \langle \O_k
\rangle_{m^2,\x} \cr
&= \lim_{\ep \to 0} \Big[

\int_{\scriptstyle |r' - r| \ge \ep \atop\scriptstyle r_0 \ge r'  \ge
\ep}
d^4 r'~ \langle (\Ox(r') - \langle \Ox \rangle)
(\O_i (r) \O_j (0) - C_{ij,k} (r) \O_k (0)) \rangle_{m^2,\x} \cr
&~~~ + ( A_{\x i,k} (\ep) C_{kj,l} (r) + A_{\x j,k} (\ep) C_{ik,l} (r)

- C_{ij,k} (r) A_{\x k,l} (\ep) ) \langle \O_l \rangle_{m^2,\x} \Big]
.\cr}}
An analogous formula is valid for the partial derivative
with respect to $m^2$.

We remark that the above variational formula for the OPE
coefficients can be used recursively to evaluate the higher
order derivatives.  Especially we can express the Taylor
coefficients
of $C_{ij,k} (r)$ with respect to $m^2, \x$ in terms of multiple
integrals of correlation functions at $m^2 = \x = 0$.
Since the integrals are done over finite
ranges, there is no infrared divergence.
The possibility of writing down variational formulas of this
kind has been mentioned in

ref.~\rwilson.

The variational formula \evaropetwo\ can be generalized
to higher-point correlation functions as follows:
\eqn\evarthree{\eqalign{
& ~~~ - \partial_\x C_{ij,k} (r;m^2,\x) \cdot \langle \O_k(0)

\O_{i_1} (r_1) ... \O_{i_n} (r_n)
\rangle_{m^2,\x} \cr
&= \lim_{\ep \to 0}

\Big[ \int_{\scriptstyle |r' - r|

\ge \ep \atop\scriptstyle r_0 \ge r'  \ge \ep} d^4 r'~

\langle (\Ox(r') - \langle \Ox \rangle)
(\O_i (r) \O_j (0) - C_{ij,k} (r) \O_k (0)) \cr
&~~~~~~~~~~~~~~~~~~~~~~~~~~~~~~~~~~~ \times
\O_{i_1} (r_1) ... \O_{i_n} (r_n)
\rangle_{m^2,\x} \cr
&~~~ + ( A_{\x i,k} (\ep) C_{kj,l} (r) + A_{\x j,k} (\ep) C_{ik,l} (r)

- C_{ij,k} (r) A_{\x k,l} (\ep) ) \langle \O_l (0)

\O_{i_1} (r_1) ... \O_{i_n} (r_n)
\rangle_{m^2,\x} \Big] ,\cr}}
where $r_0$ is arbitrary as long as $r_0$ is smaller than any
of $r_1, ..., r_n$.

Finally by applying \evarope\ to the singular part of

the OPE coefficients $C_m, C_\x$ \eope, we obtain
\eqn\evarsing{\eqalign{
& ~~~ - \partial_\x C_{si,j} (r;m^2,\x) \cdot \vevx{\O_j} \cr
&= \lim_{\ep \to 0} \Big[

{\rm Sg} \int_{\scriptstyle |r' - r| \ge \ep

\atop\scriptstyle r_0 \ge r'  \ge \ep} d^4 r'~

\vevcx{\Ox (r') (\O_s (r) \O_i (0) - C_{si,j} (r) \O_j (0))} \cr
&~~~ + ( A_{\x s,j} (\ep) C_{ij,k} (r) + A_{\x i,j} (\ep) C_{sj,k} (r)

- C_{si,j} (r) A_{\x j,k} (\ep) ) \vevx{\O_k}

\Big] ,\cr}}
where Sg stands for the singular part with respect
to $r$, and $s = m, \x$.
The purpose of this paper is to use this formula to compute
the first order corrections in the OPE coefficients
$C_m, C_\x$.

\newsec{Determination of $\Om, \Ox$ at $\x = 0$}

At $\x = 0$ we have a free massive $\phi^4$ theory
with the mass parameter $m^2$.
We consider two composite operators $\phitwo$
and $\phifour$ which are uniquely specified
by the conditions
\eqn\ephi{\vev{\phitwo} = {v \over 2},~
\vev{\phifour} = {v^2 \over 8} ,}
where
\eqn\ev{v \equiv {m^2 \over 16 \pi^2}~ \ln {m^2 \e^{2 \gamma}
\over 4} ,}
and $\gamma = 0.577...$ is the Euler constant.
(Recall that the renormalization point is taken at
$r = 1$.)

The correlation function of two $\phi$'s (i.e., propagator)

is given by
\eqn\edelta{\eqalign{\D(r;m^2) &\equiv

\vev{\phi (r) \phi (0)} \cr
&= {m K_1 (mr) \over 4 \pi^2 r} \cr
&= {1 \over 4 \pi^2 r^2} +

v + {m^2 \over 8 \pi^2}~\left( \ln r  - {1 \over 2} \right) \cr
&~~~ + r^2 \left( {1 \over 8} ~m^2 v + {m^4 \over 64 \pi^2}
\left( \ln r

- {5 \over 4} \right) \right) + O \Big( r^4 \Big) .\cr}}

At $\x = 0$ the most general scalar operator of scaling
dimension two is a linear combination of $m^2 \one$
and $\phitwo$.  Hence,
\eqn\eOm{\Om = \phitwo + a m^2 \one ,}
where the choice of the coefficient of $\phitwo$
corresponds
to a particular normalization of $m^2$.  Likewise,
we expect, at $\x =0$,
\eqn\eOx{
\Ox = \phifour + b m^2 \phitwo + {c \over 2 \pi^4} m^4 \one
+ d \partial^2 \phitwo .}

To determine the unknown coefficients $a, b, c$, and $d$
we apply
the variational formula (obtained from \evarm)
\eqn\evarvev{- \partial_{m^2} \vev{\O_i} = \lim_{\ep \to 0}
\Big[ \int_{r \ge \ep} d^4 r~ \vevc{\Om (r) \O_i (0)}
+ A_{mi,j} (\ep) \vev{\O_j} \Big] ,}
where $i = \one, m, \x, \R$.  From \eA\ we find
the subtractions $A_m$ at $\x = 0$ as
\eqn\eAmzero{A_m (\ep;m^2,0) = \left( \matrix{
0&0&0&0\cr
- 2 b_{\one,0} \ln \ep&0&0&0\cr
{1 \over \ep^2}~c_{m\x,\one} (m^2 \ep^2,0) - 2 m^2 b_{\one,1}
\ln \ep&
- b_{m,1} \ln \ep&0&0\cr
{1 \over \ep^2}~c_{m\R,\one} (m^2 \ep^2,0)

- 2 m^2 b_{\one,0} c_{m\R,m} (0) \ln \ep& c_{m\R,m}
(0)&0&0\cr} \right) .}

Using \evarvev\ for $i = m$, we find
\eqn\evarvevm{- \partial_{m^2} \vev{\Om} = - {1 \over 32
\pi^2}

- {v \over 2 m^2} - \left( {1 \over 16 \pi^2} + 2 b_{\one,0} \right)
\ln \ep ,}
where we used the integral
\eqn\eIi{\int_{r \ge \ep} d^4 r~ {1 \over 2} ~\D(r;m^2)^2 =

- {1 \over 32 \pi^2}

- {v \over 2 m^2} - {1 \over 16 \pi^2}~\ln \ep .}

On the other hand, from \ephi, \ev, and \eOm, we find
\eqn\evarvevmtwo{- \partial_{m^2} \vev{\Om} = - {1 \over 32
\pi^2}

- {v \over 2 m^2} - a .}
Hence, we obtain
\eqn\ea{a = 0}
and

\eqn\ebonezero{b_{\one,0} = - {1 \over 32 \pi^2} .}

Similarly, using \evarvev\ for $i = \R$ and

\eqn\evarvevR{ \vev{\R} = 0 }
(this is a consequence of translation invariance), we obtain
\eqn\ecmRone{c_{m\R,\one} (m^2,0) = - {1 \over 4 \pi^2}

+ {m^2 \over 8 \pi^2} ,~~ c_{m\R,m} (0) = - 2.}

Finally let us determine $\Ox$.  From \eCmCx\ and \eH\ we
find
\eqn\ecxxRzero{C_{\x\x,\R} (r;\x=0) = 0.}
This implies
\eqn\ed{ d = 0 .}
(See the next section for more on the OPE coefficients
at $\x = 0$.)
Applying \evarvev\ for $i = \x$, we find
\eqn\evarvevx{\eqalign{- \partial_{m^2} \vev{\Ox}

&= \left( {v \over 2} + b m^2 \right) \left( - {1 \over 32 \pi^2} -
{v \over 2 m^2} - {1 \over 16 \pi^2}~\ln \ep \right) \cr
&~~~ + {1 \over \ep^2}~c_{m\x,\one} (m^2 \ep^2,0)

- 2 m^2 b_{\one,1} \ln \ep - b_{m,1} {v \over 2}~\ln \ep .\cr}}
On the other hand, we find, from \ephi, \ev, and \eOx,
\eqn\evarvevxtwo{ - \partial_{m^2} \vev{\Ox} =
{v \over 2} \left( - {1 \over 32 \pi^2} - {v \over 2 m^2} \right)
- {c\over \pi^4}~m^2 .}
Hence, we obtain
\eqn\eb{ b = 0 }
and

\eqn\eboneone{b_{\one,1} = 0,~
b_{m,1} = - {1 \over 16 \pi^2} ,~
c_{m\x,\one} (m^2,0) = - {c \over \pi^4}~m^2 .}
The constant $c$ is undetermined at this stage, but
it will be fixed later in sect.~7.

\newsec{Determination of the subtractions $A_{\x}$
at $\x = 0$}

In order to apply the variational formula \evarsing\ to
calculate the first order corrections to $C_m, C_\x$, we
need to know the subtractions $A_\x (\ep;m^2,0)$
given by \eA.  (Its concrete expression is somewhat
lengthy and is given in Appendix B.)  Especially
this requires all finite counterterms at $\x = 0$.
Some beta functions and finite counterterms have

already been
determined in the previous section.  Most others, which
are necessary to evaluate $A_\x$, can be obtained from
the OPE coefficients at $\x = 0$.  Using \eCmCx\ and
\eH\ we can fit the OPE coefficients, calculated
by using the propagator \edelta, in terms of
the beta functions and finite counterterms.

Now that we have determined $\Om, \Ox$ at $\x = 0$,
we can compute the singular part of their OPE
coefficients as follows (the undetermined constant $c$ in
\eOx\ does not affect the coefficients):
\eqn\eopelist{\eqalign{C_{mm,\one} (r) &
= {1 \over 32 \pi^4 r^4} \cr
C_{m\x,\one} (r) &= 0, ~
C_{m\x,m} (r) = {1 \over 32 \pi^4 r^4} \cr
C_{m\R,\one} (r) &= {1 \over 4 \pi^4 r^6}

- {m^2 \over 16 \pi^4 r^4} ,~
C_{m\R,m} (r) = 0 \cr
C_{\x\x,\one} (r) &= {1 \over 6144 \pi^8 r^8}
+ \left(  - { 1 \over 6144} + {\ln r \over 3072} \right)
{m^2 \over \pi^8 r^6} \cr
& ~~~~~~~~~~ + \left({1 \over 98304} - {5 \ln r \over 24576}
+ {(\ln r)^2 \over 4096} \right) {m^4 \over \pi^8 r^4} \cr
C_{\x\x,m} (r) &= {1 \over 192 \pi^6 r^6} + \left(
- {5 \over 1536} + {\ln r \over 128} \right)

{m^2 \over \pi^6 r^4}\cr
C_{\x\x,\x} (r) &= {3 \over 16 \pi^4 r^4} ,~
C_{\x\x,\R} (r) = 0 \cr
C_{\x\R,\one} (r) &= 0 ,~
C_{\x\R,m} (r) = {1 \over 4 \pi^4 r^6}

- {m^2 \over 16 \pi^4 r^4} \cr
C_{\x\R,\x} (r) &= 0 ,~ C_{\x\R,\R} (r)

= {1 \over 32 \pi^4 r^4} .\cr}}

On the other hand we can express the OPE coefficients in
terms of the beta functions and finite counterterms using
\eCmCx\ and \eH.  This is done in Appendix C.  By
comparing
this result with the above coefficient functions \eopelist\

we can extract the following values of
the beta functions and finite counterterms:
\eqn\eresultone{\eqalign{b_{\one,0} &= - {1 \over 32 \pi^2},~
b_{\one,1} = 0,~ b_{\one,2} = {1 \over \pi^6} \left(
- {1 \over 49152} + { 5 c \over 16} \right) + {1 \over 16 \pi^2}~
\partial_{m^2} c_{\x\x,m} (m^2,0) \cr
b_{m,1} &= - { 1 \over 16 \pi^2} ,~
b_{m,2} = {5 \over 768 \pi^4} ,~~~
b_{\x,1} = - {3 \over 8 \pi^2} \cr

c_{m\x,\one} (m^2,0) &= - { c \over \pi^4} ~m^2 ,~~~
c_{m\R,\one} (m^2,0) =

{1 \over 8 \pi^2} (- 2 + m^2) ,~~~
c_{m\R,m} (0) = - 2 \cr
c_{\x\x,\one} (m^2,0) &= - {1 \over 12288 \pi^6} + {m^4 \over
2}~
\partial_{m^2}^2 c_{\x\x,\one} (m^2,0) \cr
c_{\x\x,m} (m^2,0) &= - {1 \over 192 \pi^4} + m^2
\partial_{m^2}
c_{\x\x,m} (m^2,0) \cr
c_{\x\R,\one} (m^2,0) &= {m^4 \over 2}~\partial_{m^2}^2

c_{\x\R,\one} (m^2,0) \cr
c_{\x\R,m} (m^2,0) &= {1 \over 8 \pi^2} (- 2 + m^2) ,~~~
c_{\x\R,\x} (0) = - 4  .\cr}}
Recall that $c_{mm,\one} (\x), c_{m\x,m} (\x)$,

and $c_{\x\x,\x} (\x)$
are all zero due to the gauge conditions \egauge.
The above results leave the following maximal
finite counterterms
\eqn\eundet{
c, ~\partial_{m^2}^2 c_{\x\x,\one} (m^2,0) ,~

\partial_{m^2} c_{\x\x,m} (m^2,0),~ c_{\x\x,\R} (0),~
\partial_{m^2}^2 c_{\x\R,\one} (m^2,0) ,~ c_{\x\R,\R} (0)}
still undetermined at $\x = 0$.

In conclusion we have determined the subtractions

$A_{\x} (\ep;m^2,0)$ as follows:
\eqn\eAresult{\eqalign{A_{mm,\one} (\ep;0) &=
{\ln \ep \over 16 \pi^2} \cr
A_{\x m,\one} (\ep;m^2,0) &= - c {m^2 \over \pi^4} \cr
A_{\x m,m} (\ep;0) &= {\ln \ep \over 16 \pi^2} \cr
A_{\x\x,\one} (\ep;m^2,0) &= - {1 \over 12288 \pi^6 \ep^4} -
{m^2 \ln \ep \over 3072 \pi^6 \ep^2} \cr
&~~~ + {m^4 \over \pi^6} \Bigg(

{\pi^6 \over 2}~\partial_{m^2}^2 c_{\x\x,\one} (m^2,0)
+ \left( {1 \over 49152} - {3 c \over 16} \right) \ln \ep \cr
&~~~~~~~~~~ - { 5 \over 24576} ~\ln^2 \ep

+ {1 \over 6144} ~\ln^3 \ep \Bigg) \cr
A_{\x\x,m} (\ep;0) &= - { 1 \over 192 \pi^4 \ep^2}

+ {m^2 \over \pi^4}
\left( \pi^4 \partial_{m^2} c_{\x\x,m} (m^2,0)

- {5 \over 768} ~\ln \ep
+ {1 \over 128} ~\ln^2 \ep \right) \cr
A_{\x\x,\x} (\ep;0) &= {3 \ln \ep \over 8 \pi^2} ,~~
A_{\x\x,\R} (\ep;0) = c_{\x\x,\R} (0) . \cr}}
Here we did not give $A_{\x\R,i}$, since we will not calculate
the first order corrections of the operator products

involving $\R$ in sect.~8.

\newsec{Curvature}

In order to obtain further information on the finite
counterterms, let us examine the commutativity
condition \ecurv\ at $\x = 0$, which gives
\eqn\efirst{\eqalign{
\big(\Omega_{\x m} (m^2,0) \big)_{m,\one} &= {c \over \pi^4}
\cr
\big(\Omega_{\x m} (m^2,0) \big)_{\x,\one} &=

\partial_\x c_{m\x,\one} (m^2,0) - {1 \over 4 \pi^2}~
c_{\x\x,\R} (0) \cr
&~~~~~~~~~~+ m^2 \left( {1 \over 8 \pi^2}~
c_{\x\x,\R} (0) - \partial_{m^2}^2 c_{\x\x,\one} (m^2,0)

\right) \cr
\big(\Omega_{\x m} (m^2,0) \big)_{\x,m} &= - 2 c_{\x\x,\R} (0)
- \partial_{m^2} c_{\x\x,m} (m^2,0) .\cr}}
We do not consider the $(\R,\one)$, $(\R,m)$ elements,
since we do not need them in the next section.

Now the curvature is defined by (see \eOmega)
\eqn\eOmegazero{\eqalign{& \Omega_{\x m} (m^2,0)
\vev{\O} =
\int_{1 \ge r} d^4 r~ {\rm F.P.}
\int_{1 \ge r'} d^4 r'~\cr
&~~~~~ \vevc{ \phitwo (r) (\phifour (r') \O(0) - C_\x (r') \O(0))

- \phifour (r) (\phitwo (r') \O(0) - C_m (r') \O(0)) }~, \cr}}
where F.P. stands for the finite part that can
be obtained by subtracting terms of order $1/r^4$
or more singular.
The calculation is straightforward, and we sketch
it in Appendix D.  The result is
\eqn\esecond{\eqalign{
\big(\Omega_{\x m} (m^2,0) \big)_{m,\one} &=

{1 \over 1024 \pi^4} \cr
\big(\Omega_{\x m} (m^2,0) \big)_{\x,\one} &=

- {1 \over 12288 \pi^6} - {5 m^2 \over 32768 \pi^6} \cr
\big(\Omega_{\x m} (m^2,0) \big)_{\x,m} &=

- {1 \over 512 \pi^4} .\cr}}

By imposing the equality between \efirst\ and \esecond,
we obtain
\eqn\eresulttwo{\eqalign{
&c = {1 \over 1024} \cr
&\partial_\x c_{m\x,\one} (0,0) - {1 \over 4 \pi^2}~
c_{\x\x,R} (0) = - {1 \over 12288 \pi^6} \cr
&\partial_\x \partial_{m^2} c_{m\x,\one} (0,0) -
\partial_{m^2}^2 c_{\x\x,\one} (0,0) + {1 \over 8 \pi^2}~
c_{\x\x,\R} (0) = - {5 \over 32768 \pi^6} \cr
&c_{\x\x,\R} (0) + {1 \over 2}~\partial_{m^2} c_{\x\x,m} (0,0) =
{1 \over 1024 \pi^4} .\cr}}

\newsec{OPE coefficients to first order}

In this section we apply \evarsing\ to calculate the
first order corrections to the operator product coefficient
functions $C_m , C_\x$.   As a final preparation
let us first introduce the integrals
\eqn\ekmsdef{\eqalign{k_{m,m\x} (r;\ep) &\equiv {\rm Sg}
\int_{\scriptstyle |r' - r|

\ge \ep \atop\scriptstyle r_0 \ge r'  \ge \ep}
d^4 r'~ \vevc{\phitwo (r') \left( \phitwo (r) \phifour (0) -
C_{m\x,m} (r) \phitwo (0) \right)} \cr
k_{m,\x\x} (r;\ep) &\equiv {\rm Sg}
\int_{\scriptstyle |r' - r|

\ge \ep \atop\scriptstyle r_0 \ge r'  \ge \ep}

d^4 r'~ \langle \phitwo (r') \cr
&~~~ \times \left( \phifour (r) \phifour (0) - C_{\x\x,m} (r)
\phitwo (0)

- C_{\x\x,\x} (r) \phifour (0) \right) \rangle_{m^2,0}^c ,\cr}}
where Sg stands for an unintegrable part with respect to $r$.
The above integrals are independent of $r_0$; in the
following
we will take it as infinity.   A variational formula

that describes the derivative of a coefficient function with
respect
to $m^2$ can be obtained from \evarsing\ by replacing $\x$
by $m^2$,
and it gives
\eqn\evarmsing{\eqalign{
&- \partial_{m^2} C_{m\x,\one} (r) = \lim_{\ep \to 0} \Big(
k_{m,m\x} (r;\ep) + A_{m\x,m} (\ep) C_{mm,\one} (r)
- C_{m\x,m} (r) A_{mm,\one} (\ep) \Big) \cr
& - \partial_{m^2} C_{\x\x,\one} (r)

- \partial_{m^2} C_{\x\x,m} (r) \cdot {v \over 2} \cr
&~~~ = \lim_{\ep \to 0} \Big[
k_{m,\x\x} (r;\ep) + 2 A_{m\x,m} (\ep) \left( C_{m\x,m} (r) {v
\over 2}
+ C_{m\x,\one} (r) \right) \cr
&~~~~~~~~~~ - C_{\x\x,\x} (r) \left(A_{m\x,\one} (\ep)

+ A_{m\x,m} (\ep) {v \over 2} \right)
- C_{\x\x,m} (r) A_{mm,\one} (\ep) \Big]  .\cr}}
By substituting the coefficient functions \eopelist\
and subtractions \eAresult\ into the above,
we obtain
\eqn\ekms{\eqalign{k_{m,m\x} (r;\ep) &= 0 \cr
k_{m,\x\x} (r;\ep) &= {1 \over \pi^8 r^6}~\left(
{1 \over 6144} + {1 \over 3072}~(\ln \ep - \ln r) \right) \cr
&~~ + {1 \over \pi^8 r^4} \Bigg( \pi^2 v \left(
{5 \over 3072} + {1 \over 256}~(\ln \ep - \ln r) \right) \cr
&~~~~~ + m^2 \left( - {1 \over 49152} - { 5 \ln \ep \over 24576}
+ {5 \ln r \over 12288} + {\ln r \over 2048} ~(\ln \ep - \ln r)
\right) \Bigg) .\cr}}

We now define
\eqn\ekxsdef{\eqalign{k_{\x,mm} (r;\ep) &\equiv

{\rm Sg} \int_{\scriptstyle |r' - r| \ge \ep

\atop\scriptstyle r'  \ge \ep} d^4 r'~
\vevc{\phifour (r') \phitwo (r) \phitwo (0)} \cr
k_{\x,m\x} (r;\ep) &\equiv
{\rm Sg} \int_{\scriptstyle |r' - r| \ge \ep

\atop\scriptstyle r'  \ge \ep} d^4 r'~\vevc{\phifour (r')
\left( \phitwo (r) \phifour (0) - C_{m\x,m} (r) \phitwo (0) \right)}
\cr
k_{\x,\x\x} (r;\ep) &\equiv

{\rm Sg} \int_{\scriptstyle |r' - r| \ge \ep

\atop\scriptstyle r'  \ge \ep} d^4 r'~
\langle \phifour (r') \cr
&~~~~~ \times \left( \phifour (r) \phifour (0) -
C_{\x\x,m} (r) \phitwo (0) - C_{\x\x,\x} (r) \phifour (0) \right)
\rangle_{m^2,0}^c .\cr}}
Then the variational formula \evarsing\ gives
\eqn\edcmm{- \partial_\x C_{mm,\one} (r;0) = \lim_{\ep \to 0}
\Big[ k_{\x,mm} (r;\ep) + 2 A_{m\x,m} (\ep) C_{mm,\one} (r)
\Big] }
\eqn\edcmx{\eqalign{& - \partial_\x C_{m\x,m} (r;0) \cdot {v
\over 2}
- \partial_\x C_{m\x,\one} (r;m^2,0) = \lim_{\ep \to 0} \Bigg[
k_{\x,m\x} (r;\ep) \cr
&~~~~~~~~~~ + (A_{m\x,m} (\ep) + A_{\x\x,\x} (\ep))  ~{\rm
Sg}
\vevc{\phitwo (r) \phifour (0)}
+ A_{\x\x,\R} (\ep) ~{\rm Sg} \vevc{\phitwo (r) \partial^2
\phitwo (0)} \cr
&~~~~~~~~~~ - C_{m\x,m} (r) \left( A_{m\x,m} (\ep) {v \over
2}
+ A_{m\x,\one} (\ep) \right) \Bigg] \cr}}
\eqn\edcxx{\eqalign{&- \partial_\x C_{\x\x,\x} (r;0) \cdot
\left({v^2 \over 8} + c {m^4 \over 2 \pi^4} \right)

- \partial_\x C_{\x\x,m} (r;m^2,0) \cdot {v \over 2}
- \partial_\x C_{\x\x,\one} (r;m^2,0) \cr
&~~~ = \lim_{\ep \to 0} \Bigg[ k_{\x,\x\x} (r;\ep)

+ 2 A_{\x\x,\x} (\ep) ~{\rm Sg} \vevc{\phifour (r) \phifour (0)}
\cr
&~~~~~ + 2 A_{\x\x,\R} (\ep) ~{\rm Sg} \vevc{\phifour (r)
\partial^2 \phitwo (0)} + 2 A_{\x\x,m} (\ep) ~{\rm Sg}
\vevc{\phitwo (r) \phifour (0)} \cr
&~~~~~ - C_{\x\x,\x} (r) \left( A_{\x\x,\x} (\ep) \left(
{v^2 \over 8} + c {m^4 \over 2 \pi^4} \right) + A_{\x\x,m} (\ep)
{v \over 2} + A_{\x\x,\one} (\ep) \right) \cr
&~~~~~ - C_{\x\x,m} (r) \left( A_{m\x,m} (\ep) {v \over 2} +
A_{m\x,\one}
(\ep) \right) \Bigg] .}}

We sketch the calculations of the integrals $k_{\x,mm}$,
$k_{\x,m\x}$, and $k_{\x,\x\x}$ in Appendix E.
(The integrals $k_{m,m\x}$ and $k_{m,\x\x}$ are
necessary for the calculations.)  By substituting
the results into the above \edcmm, \edcmx, and \edcxx,
we can find the first order corrections to the OPE coefficient
functions.  But we must recall that the coefficient functions
are completely determined in terms of the beta functions
($\b1$, $\bm$, and $\bx$) and finite counterterms
($c_m$ and $c_\x$), and the relation is given by
eqs.~\eCmCx\
and \eH.  Hence we have an indirect way of
computing $k_{\x,mm}$, $k_{\x,m\x}$, and $k_{\x,\x\x}$
in terms of the beta functions and finite counterterms;
namely we first calculate the derivatives of
the coefficient functions in terms of the beta functions
and finite counterterms, and then
we substitute the results into the above variational
formulas \edcmm, \edcmx, and \edcxx.  This is done
in Appendix F.  By comparing the two results for
the integrals, we can obtain the following beta functions
and finite counterterms:
\eqn\eresultthree{\eqalign{
b_{\one,3} &= {11 \over \pi^8}~\left({1 \over 131072} - {c \over
128}\right)
- {1 \over 8 \pi^2} (\partial_\x \partial_{m^2} c_{m\x,\one} (0,0)
- \partial_{m^2}^2 c_{\x\x,\one}(0,0))  \cr
&~~~~~ + {1 \over 16 \pi^2} \partial_\x \partial_{m^2}
c_{\x\x,m} (0,0)
- {3 \over 4 \pi^2}~\partial_{m^2}^2 c_{\x\x,\one} (0,0) \cr
b_{m,3} &= - {29 \over 8192 \pi^6} + {c \over 8 \pi^6} ,~~

b_{\x,2} = {17 \over 128 \pi^4} \cr
\partial_{m^2} c_{\x\x,m} (0,0) &= 0 ,~~
c_{\x\x,\R} (0) = {3 \over 2048 \pi^4} - {c \over 2 \pi^4}  .\cr
\partial_\x c_{m\x,\one} (0,0) &= {1 \over \pi^6} \left(

{7 \over 24576 } - {c \over 8} \right) \cr
\partial_{\x} c_{\x\x,\one} (0,0) &= {1 \over 65536 \pi^8} ,~~
\partial_\x \partial_{m^2} c_{\x\x,\one} (0,0) = {1 \over \pi^8}~
\left( {1 \over 16384} + {c \over 192} \right) \cr
\partial_\x c_{\x\x,m} (0,0) &= {1 \over \pi^6} \left( {43 \over
12288}

- {c \over 4} \right) .\cr}}

Now we combine the above results with the results

\eresulttwo\ of the previous section, and we obtain

\eqn\eresultfour{\eqalign{
b_{\one,3} &=  {9 \over 262144 \pi^8}

+ {1 \over 16 \pi^2} \partial_\x \partial_{m^2} c_{\x\x,m} (0,0)
- {3 \over 4 \pi^2}~\partial_{m^2}^2 c_{\x\x,\one} (0,0) \cr
b_{m,3} &= - {7 \over 2048 \pi^6} ,~~

b_{\x,2} = {17 \over 128 \pi^4} \cr
\partial_{m^2} c_{\x\x,m} (0,0) &= 0 ,~~
c_{\x\x,\R} (0) = {1 \over 1024 \pi^4}  \cr
\partial_\x c_{m\x,\one} (0,0) &= {1 \over 6144 \pi^6} ,~~
\partial_\x \partial_{m^2} c_{m\x,\one} (0,0) = \partial_{m^2}^2

c_{\x\x,\one} (0,0) + {9 \over 32768 \pi^6} \cr
\partial_{\x} c_{\x\x,\one} (0,0) &= {1 \over 65536 \pi^8} ,~~
\partial_\x \partial_{m^2} c_{\x\x,\one} (0,0) = {13 \over
196608 \pi^8} \cr
\partial_\x c_{\x\x,m} (0,0) &= {5 \over 1536 \pi^6} .\cr}}
Thus, all the counterterms at $\x = 0$, except for

$\partial_{m^2}^2 c_{\x\x,\one} (0,0)$,
are determined.\foot{We can find
$\partial_{m^2}^2 c_{\x\R,\one} (0,0) = 1/(256 \pi^4)$
and $c_{\x\R,\R} (0) = - 3/(64 \pi^2)$ from a separate
consideration.
These are unnecessary for the subtractions \eAresult.}
The unknown maximal finite counterterm
leaves the operator $\Ox$ undetermined by
\eqn\eamb{- {m^4 \over 2}~\partial_{m^2}^2 c_{\x\x,\one}
(0,0) \one .}
We need the next order calculation to remove this
ambiguity.

In conclusion we obtain, using \eresultfour\
and the integrals evaluated in Appendix E,
the following first order corrections
to the OPE coefficient functions:
\eqn\edxCmmone{\partial_\x C_{mm,\one} (r) = - {\ln r \over
256 \pi^6 r^4}}
\eqn\edxCmxone{\partial_\x C_{m\x,\one} (r) =  {1 \over
\pi^8}
\left( - {1 \over 6144 r^6} + {m^2 \over r^4} \left( - {1 \over
24576}
- {\ln^2 r \over 4096} \right) \right) }
\eqn\edxCmxm{\partial_\x C_{m\x,m} (r) = {1 \over \pi^6 r^4}
\left(
- {5 \over 1536} - {3 \ln r \over 256} \right)}
\eqn\edxCxxone{\eqalign{\partial_\x C_{\x\x,\one} (r) &=

{1 \over \pi^{10}} \Bigg[
- {\ln r \over 8192 r^8} + {m^2 \over r^6}
\left( {5 \over 196608} + {\ln r \over 16384} - {\ln^2 r \over
4096} \right) \cr
&~~~ + {m^4 \over r^4} \left(
- {5 \over 786432} + {3 \pi^6 \over 32} ~
\partial_{m^2}^2 c_{\x\x,\one} (0,0) + {17 \ln r \over 131072} +
{3 \ln^2 r \over 131072} - {5 \ln^3 r \over 32768} \right)

\Bigg] \cr}}
\eqn\edxCxxm{\partial_\x C_{\x\x,m} (r) =
{1 \over \pi^8} \Bigg[

{1 \over r^6} \left( - {3 \over 2048} - {11 \ln r \over 3072} \right)
+ {m^2 \over r^4} \left(
{7 \over 4096} - {13 \ln r \over 6144} - {9 \ln^2 r \over 2048}
\right) \Bigg] }
\eqn\edxCxxxR{\partial_\x C_{\x\x,\x} (r) =
{1 \over \pi^6 r^4} \left(
- {17 \over 256} - {9 \ln r \over 128} \right) ,~~
\partial_\x C_{\x\x,\R} (r) = - {11 \over 32768 \pi^8 r^4} .}
The unknown term in \edxCxxone\ is due to the unknown

first order correction \eamb\ to $\Ox$.

\newsec{Conclusions}

In this paper we have derived the variational formula
\evarsing\
for OPE coefficients from the general variational formula
\evarx\ for
correlation functions.  We have applied \evarsing\ to
calculate the first
order corrections to the OPE coefficients; the results are
given by eqs.~\edxCmmone\ to \edxCxxxR.
The OPE coefficients $C_m$ and $C_\x$ (defined by
\eope) are determined uniquely in terms of the beta functions
and
finite counterterms as given by \eCmCx\ and \eH, while the
finite
counterterms are constrained by the commutativity
condition
\ecurv.  In principle we can perform higher order calculations
of the OPE coefficients by using the variational formula
\evarsing\
recursively together with the relations \eCmCx, \eH, and the
condition \ecurv.

Under the gauge condition \egauge\ we
have the relations \ethree\ that determine
the OPE coefficients $C_{mm,\one}$, $C_{m\x,m}$,
$C_{\x\x,\x}$ uniquely in terms of
the beta functions $\b1$, $\bm$,
$\bx$; using the OPE coefficients that we have calculated,
we obtain
\eqn\ebetaresult{\eqalign{
\b1 (\x) &= - {1 \over 32 \pi^2}  + {7 \over 49152 \pi^6}~\x^2 +
O (\x^3) \cr
\bm (\x) &= - {1 \over 16 \pi^2}~\x + {5 \over 1536 \pi^4}~\x^2

- {7 \over 12288 \pi^6}~\x^3 + O(\x^4) \cr
\bx (\x) &= - {3 \over 16 \pi^2}~\x^2 + {17 \over 768
\pi^4}~\x^3

+ O(\x^4) . \cr}}
(Recall that our convention differs from the standard
one by sign.)
Up to this order the beta function $\bx$ is scheme
independent,
and it agrees with the known result.
On the other hand only the first order term
of $\bm$ is scheme independent.  Up to the third order
that we calculated, our result agrees with that in the
dimensional regularization with the minimal subtraction.
See, for example, ref.~\ref\rvladimirov{A. A. Vladimirov,
D. I. Kazakov, and O. V. Tarasov, Sov. Phys. JETP
50 (1979)521}.
(I could not find any previous calculation of $\b1$ to
compare.)
In our scheme
the condition \egauge\ amounts to taking certain finite

subtractions as zero in the variational formulas; it is
analogous to the minimal subtraction which requires
subtraction of only the divergent parts.
We speculate that our choice of $m^2$ and $\x$ given by
the gauge condition \egauge\ is equivalent to the
dimensional
regularization with the minimal subtraction.

We finally mention that renormalization in the coordinate
space has recently been considered in ref.~\ref\rfreedman{
D. Z. Freedman, K. Johnson, and J. I. Latorre,
``Differential regularization and renormalization:
a new method of calculation in quantum field theory'',
preprint CTP\#1972 (May 1991)}
and found
to be a useful tool for perturbative calculations.
As the authors themselves explain in sect.~II.A
of their paper, their method (called the differential
renormalization) is equivalent to regulating
divergent integrals by excluding infinitesimal balls
and adding counterterms, which is exactly what we have
done
in this paper.  Hence, the differential renormalization scheme
is more or less the same as ours, except that it
corresponds to
a different choice of parameters $\x, m^2$.  The authors
of \rfreedman\ have shown that the differential
renormalization
is not equivalent (but physically equivalent) to the
dimensional
regularization with the minimal subtraction.

\vfill
\def\Gtwo{G^{(2)} (p^2;m^2,\x)}
\def\Gtwom{G^{(2)}_m (p^2;m^2,\x)}
\def\Gtwomph{G^{(2)}_{m_{ph}} (p^2;m^2,\x)}
\appendix{A}{Mass insertion in Callan-Symanzik equations}

In this section we explain the relation between the
mass-independent
RG equations adopted in this paper and the
Callan-Symanzik

equations \rcseqs.  Especially we explain the relation
between
the mass insertions in the two approaches.

The elementary scalar operator $\phi$ satisfies the RG
equation
\eqn\eRGphi{{d \over dl}~\phi (r) = (1 + \gamma (\x)) \phi (r) .}
We define the Fourier transforms $G^{(2)} (p^2;m^2,\x)$ and
$G^{(2)}_m (p^2;m^2,\x)$ by
\eqn\eFourier{\eqalign{G^{(2)} (p^2;m^2,\x) &\equiv
\int d^4 r~\e^{- i p r} \vevcx{\phi (r) \phi (0)} \cr
G^{(2)}_m (p^2;m^2,\x) &\equiv \int d^4 r ~
\e^{- i p r} \int d^4 r'~ \vevcx{\Om (r') \phi (r) \phi (0)} .\cr}}
Note that the operator $\Om$ has only dimension two, and
the integral over $r'$ does not require any subtraction.
The variational formula \evarm\ implies
\eqn\emass{- \left( {\partial \over \partial m^2} \right)_\x
G^{(2)} (p^2;m^2,\x) = G^{(2)}_m (p^2;m^2,\x) .}

Now the RG eq. \eRGphi\ implies
\eqn\eRGGtwo{\left( 2 p^2 {\partial \over \partial p^2}
+ (2 + \bm) {\partial \over \partial m^2} + \bx {\partial \over

\partial_\x} \right) \Gtwo = (-2 + 2 \gamma) \Gtwo .}
The Callan-Symanzik equation can be obtained by
rewriting this RG eq. in terms of the physical mass
$m_{ph}^2$ and the physical coupling constant $u$.
The physical mass is defined from the small $p^2$
behavior
\eqn\emph{\Gtwo = {z(m^2,\x) \over m_{ph}^2 + p^2 +
o(p^2)} .}
Let $\Gamma^{(4)} (p_1,p_2,p_3;m^2,\x)$ be the one-particle
irreducible four-point function in the momentum space.
Then,
the physical coupling is defined by
\eqn\eu{\Gamma^{(4)} (0,0,0;m^2,\x) = {u \over z(m^2,\x)^2} ,}
where $z(m^2,\x)$ is defined by \emph.  The physical
coupling
is invariant under the RG.

The mass insertion that will appear in the Callan-Symmanzik
equation
differs from the mass insertion, defined by \eFourier, by
a factor:
\eqn\einsert{\Gtwomph \equiv \left( {\partial m^2 \over \partial
m_{ph}^2}
\right)_\x \Gtwom .}
Then, \emass\ implies

\eqn\emassph{\Gtwomph = \left( {\partial \over \partial
m_{ph}^2} \right)_\x
\Gtwo .}

If we renormalize the Fourier transforms by
\eqn\eFouriertwo{\eqalign{G^{(2)R} (p^2;m_{ph}^2,u) &\equiv

{1 \over z(m^2,\x)}~\Gtwo \cr
G^{(2)R}_{m_{ph}} (p^2;m_{ph}^2,u) &\equiv {1 \over
z(m^2,\x)}~\Gtwom ,\cr}}
then the Callan-Symanzik equation can be obtained from
\eRGGtwo\ as
\eqn\ecs{\left( 2 p^2 {\partial \over \partial p^2} + b(u) \left(
{\partial
\over \partial u} \right)_{m_{ph}^2} + 2 (1-C(u)) \right)
G^{(2)R} (p^2;m_{ph}^2,u) = 2 m_{ph}^2 G^{(2)R}_{m_{ph}}

(p^2;m_{ph}^2,u) ,}
where the RG invariant functions $b(u)$ and $C(u)$ are
defined by
\eqn\ebC{\eqalign{b(u) &\equiv - 2 m_{ph}^2 \left( {\partial u
\over \partial
m_{ph}^2} \right)_\x \cr
C(u) &\equiv \gamma (\x) - {1 \over 2}~b(u) \left(
{\partial \ln z(m^2,\x) \over \partial u} \right)_{m_{ph}^2} .\cr}}
See ref.~\ref\rparisi{G. Parisi, {\it Statistical Field Theory}
(Addison-Wesley,
Menlo Park, 1988)} for more details.

\filbreak

\appendix{B}{Expressions of $A_\x (\ep;m^2,0)$ in terms
of beta functions and finite counterterms}

Using \eA\ we find
\eqn\eAxzero{\eqalign{
A_{\x\x,\one} (\ep;m^2,0) &=

{1 \over \ep^4}~c_{\x\x,\one} (m^2 \ep^2,0) \cr
&~~~ + {2 \ln \ep \over \ep^2}~

(b_{m,1} c_{m\x,\one} (m^2 \ep^2,0)
- b_{\one,0} c_{\x\x,m} (m^2 \ep^2,0)) \cr
&~~~ + m^4 \Bigg( - b_{\one,2} \ln \ep +
\left( b_{m,2} b_{\one,0} - 2 b_{m,1} b_{\one,1}

+ {1 \over 2} ~b_{\x,1} b_{\one,1} \right) \ln^2 \ep \cr

&~~~~~ + \left( {2 \over 3} b_{m,1}^2 b_{\one,0}

- {1 \over 3} b_{m,1} b_{\x,1} b_{\one,0} \right)

\ln^3 \ep \Bigg) \cr
A_{\x\x,m} (\ep;m^2,0) &= {1 \over \ep^2} ~
c_{\x\x,m} (m^2 \ep^2,0)
+ m^2 \left( - b_{m,2} \ln \ep + b_{m,1}

\left( - b_{m,1} + {1 \over 2} b_{\x,1}
\right) \ln^2 \ep \right) \cr
A_{\x\x,\x} (\ep;0) &= - b_{\x,1} \ln \ep \cr
A_{\x\x,\R} (\ep;0) &= c_{\x\x,\R} (0) ~.\cr}}
We do not list $A_{\x\R,i}$, since we do not use them
in the first order calculations of sect.~8.
Note that $A_{\x m,i}$ can be obtained from $A_{m \x,i}$

by symmetry
\eqn\eamxsymm{A_{\x m,i} (\ep;m^2,\x) = A_{m \x,i}
(\ep;m^2,\x) .}

\appendix{C}{Expressions of $C_m (r;m^2,0)$,
and $C_\x (r;m^2,0)$ in terms
of beta functions and finite counterterms}

Using \eCmCx\ and \eH\ we find the following expressions
for the OPE coefficients at $\x = 0$:
\eqn\eCmexp{\eqalign{C_{mm,\one} (r) &= {1 \over \pi^2}~
{ - b_{\one,0} \over r^4} \cr
C_{m\x,\one} (r) &= {1 \over \pi^2}  \left( - {1 \over r^6}

c_{m\x,\one} (0,0) - {m^2 \over r^4} b_{\one,1} \right) ,~~
C_{m\x,m} (r) = - {b_{m,1} \over 2 \pi^2 r^4} \cr
C_{m\R,\one} (r) &= {1 \over \pi^2} \left(
- {1 \over r^6} ~c_{m\R,\one} (0,0)
- { m^2 \over r^4} ~b_{\one,0} c_{m\R,m} (0) \right) ,~~
C_{m\R,m} (r) = 0 \cr }}
and
\eqn\eCxexp{\eqalign{
C_{\x\x,\one} (r) &= {1 \over \pi^2} \Bigg(
- {2 c_{\x\x,\one} (0,0) \over r^8} ~
+ {m^2 \over r^6}~ \Big(  2 ( - b_{m,1} c_{m\x,\one} (0,0)
+ b_{\one,0} c_{\x\x,m} (0,0) ) \ln r \cr
&~~~~~~~~ + b_{m,1} c_{m\x,\one} (0,0)
- b_{\one,0} c_{\x\x,m} (0,0)

- \partial_{m^2} c_{\x\x,\one} (0,0) \Big) \cr
&~~~ + {m^4 \over r^4} \Big( b_{m,1} b_{\one,0} (b_{m,1}

- {1 \over 2} ~
b_{\x,1} ) \ln^2 r + ( b_{m,2} b_{\one,0} - 2 b_{m,1} b_{\one,1}
+ {1 \over 2}~ b_{\x,1} b_{\one,1} ) \ln r \cr
&~~~~~~~~  - {1 \over 2}~ b_{\one,2} + b_{m,1} \partial_{m^2}

c_{m\x,\one} (0,0) - b_{\one,0} \partial_{m^2} c_{\x\x,m} (0,0)

\Big) \Bigg)\cr
C_{\x\x,m} (r) &= {1 \over \pi^2} \left(
- {c_{\x\x,m} (0,0) \over r^6} + {m^2 \over r^4}~
\left( b_{m,1} ( - b_{m,1} + {1 \over 2}~ b_{\x,1} ) \ln r

- {1 \over 2}~ b_{m,2}
\right) \right) \cr
C_{\x\x,\x} (r) &= - {b_{\x,1} \over 2 \pi^2 r^4} ,~~~

C_{\x\x,\R} (r) = 0 \cr
C_{\x\R,\one} (r) &= {1 \over \pi^2} \Bigg(
- {2 c_{\x\R,\one} (0,0) \over r^8} ~
+ {m^2 \over r^6}~ \Big( ( - b_{m,1} c_{m\R,\one} (0,0)
+ 2 b_{\one,0} c_{\x\R,m} (0,0) ) \ln r \cr

&~~~~~~~~ + {1 \over 2}~ b_{m,1} c_{m\R,\one} (0,0)

- b_{\one,0} c_{\x\R,m} (0,0)
- \partial_{m^2} c_{\x\R,\one} (0,0) \Big) \cr
&~~~ + {m^4 \over r^4} \Big(
b_{m,1} b_{\one,0}

( - 2 c_{m\R,m} (0) + c_{\x\R,\x} (0) ) \ln r \cr
&~~~~~~~~ - {1 \over 2} ~ b_{\one,1} c_{\x\R,\x} (0)
+ {1 \over 2}~

b_{m,1} \partial_{m^2} c_{m\R,\one} (0,0)

- b_{\one,0} \partial_{m^2}
c_{\x\R,m}(0,0) \Big) \Bigg) \cr
C_{\x\R,m} (r) &= {1 \over \pi^2} \left(
- {c_{\x\R,m} (0,0) \over r^6} +
{m^2 \over r^4}~ {b_{m,1} \over 2}~

(c_{m\R,m} (0) - c_{\x\R,\x} (0)) \right) \cr
C_{\x\R,\x} (r) &= 0 ,~~~ C_{\x\R,\R} (r)
 = - {b_{m,1} \over 2 \pi^2 r^4} . \cr}}

\appendix{D}{Calculation of the curvature}

\subsec{The ($m,\one$)-element}

{}From \eOmegazero\ we find
\eqn\emone{\eqalign{& (\Omega_{\x m})_{m,\one}

= \int_{1 \ge r} d^4 r~ {\rm F.P.} \int_{1 \ge r'} d^4 r' \cr
&~~~~~ \vevc{\phitwo (r) \left( \phifour (r') \phitwo (0)
- C_{\x m,m} (r') \phitwo (0) \right) -
\phifour (r) \phitwo (r') \phitwo (0)} \cr
&~~~ = \int_{1 \ge r} d^4 r~ {\rm F.P.} \int_{1 \ge r'}
d^4 r'~\Big[ {1 \over 4}~ \D(r-r')^2 (\D(r')^2 - \D(r)^2)
- {1 \over 64 \pi^4 r'^4}~ \D(r)^2 \Big] .\cr}}
We expect this to be a constant.  Hence, we can take
$m^2 = 0$ to evaluate the integrand.

For $m^2 = 0$ we find
\eqn\emonetwo{\eqalign{&{\rm F.P.} \int_{1 \ge r'}
d^4 r'~\Big[ {1 \over 4}~ \D(r-r')^2 (\D(r')^2 - \D(r)^2)
- {1 \over 64 \pi^4 r'^4}~ \D(r)^2 \Big] \cr
& ~~~ = {\rm F.P.} \int_{1 \ge r'} d^4 r'~{1 \over 1024 \pi^8}
\Big[ {1 \over |r-r'|^4}~\left( {1 \over r'^4} - {1 \over r^4}
\right) - {1 \over r^4 r'^4} \Big] \cr
& ~~~ = {1 \over 1024 \pi^6 r^2} .\cr}}

Hence, we obtain
\eqn\emoneresult{(\Omega_{\x m})_{m,\one} =

{1 \over 1024 \pi^4} ~.}

\subsec{The ($\x,m$)- and ($\x,\one$)-elements}

{}From \eOmegazero\ we obtain
\eqn\exmxone{\eqalign{&(\Omega_{\x m})_{\x,m} {v \over 2}
+ (\Omega_{\x m})_{\x,\one} \cr
&~~ = \int_{1 \ge r} d^4r~{\rm F.P.} \int_{1 \ge r'}
d^4 r' \cr
&~~~~~~~~~~\langle \phitwo (r) \left( \phifour (r')

\phifour (0) - C_{\x\x,m} (r') \phitwo (0) - C_{\x\x,\x}
(r') \phifour (0) \right) \cr
&~~~~~~~~~~~~ - \phifour (r) \left(
\phitwo (r') \phifour (0) - C_{m\x,m} (r') \phitwo (0)

\right) \rangle_{m^2,0}^c \cr
&~~ = \int_{1 \ge r} d^4r~{\rm F.P.} \int_{1 \ge r'}
d^4 r' ~\Bigg[
{1 \over 6} (\D(r')^3 - \D(r)^3) \D(r-r') \D(r') \cr
&~~~~~~~~~~ + {v \over 8}~\D(r-r')^2 (\D(r')^2 - \D(r)^2) \cr
&~~~~~~~~~~ + ( - C_{\x\x,m} (r') - v C_{\x\x,\x} (r')
+ v C_{m\x,m} (r') ) {1 \over 2}~ \D(r)^2 \Big] \cr
&~~ = \int_{1 \ge r} d^4r~{\rm F.P.} \int_{1 \ge r'}
d^4 r' ~{1 \over \pi^{10}}~\Bigg[
{\pi^2 v \over 2048}~{1 \over |r-r'|^4}~
\left( {1 \over r'^4} - {1 \over r^4} \right) \cr
&~~~~~ + \left( - {m^2 \over 8192} + {\pi^2 v \over
512} \right)~{1 \over r^2 |r-r'|^2 r'^4} +
{m^2 \over 4096}~{\ln r' \over r^2 |r-r'|^2 r'^4} \cr
&~~~~~ + \left( - {1 \over 6144 r^4} +
{m^2 \over 24576 r^2} - {\pi^2 v \over 1536 r^2} -
{m^2 \ln r \over 6144 r^2} \right)~{1 \over r'^6} \cr
&~~~~~ + \left( - {m^2 \over 24576} +
{\pi^2 v \over 1536} + {1 \over 6144 r^2} + {m^2 \ln r
\over 12288} \right)~{1 \over |r-r'|^2 r'^6} \cr
&~~~~~ + {m^2 \over 12288}~{\ln |r-r'| \over
r^2 r'^6} \Bigg] ,\cr}}
where we have expanded the integrand up to first order
in $m^2$ ($v$ is counted as first order), and we have kept
only the terms that contribute to the integrable
part with respect to $r$.

It is straightforward to find
\eqn\exmxonetwo{\eqalign{&(\Omega_{\x m})_{\x,m} {v \over
2}
+ (\Omega_{\x m})_{\x,\one} \cr
&~~= \int_{1 \ge r} d^4 r~
{1 \over \pi^8}~\Bigg[ - {1 \over 12288 r^2}

+ m^2 \left(
{1 \over 98304} - {1 \over 12288 r^2} -
{\ln r \over 24576} + {\ln r \over 6144 r^2} \right) \cr
&~~~~~ + \pi^2 v \left( - {1 \over 3072} - {5 \over
6144 r^2} \right) \Bigg] \cr
&~~= - {1 \over 12288 \pi^6} - {5 m^2 \over 32768 \pi^6}
- {v \over 1024 \pi^4} ~.\cr}}
Therefore, we obtain
\eqn\exmxoneresult{\eqalign{
(\Omega_{\x m})_{\x,m} &= - {1 \over 512 \pi^4} \cr
(\Omega_{\x m})_{\x,\one} &= - {1 \over 12288 \pi^6}

- {5 m^2 \over 32768 \pi^6} .\cr}}

\appendix{E}{Calculation of the integrals $k_{\x,mm}$,
$k_{\x,m\x}$, and $k_{\x,\x\x}$}

\subsec{$k_{\x,mm} (r;\ep)$}

This integral is independent of $m^2$ from
the dimensional analysis; it has dimension four, but we need
four
for the singularity $1/r^4$.  Hence, we can set $m^2 = 0$.
Then, we find
\eqn\ekxmmcal{\eqalign{&k_{\x,mm} (r;\ep) \equiv {\rm Sg}
\int_{\scriptstyle |r' - r| \ge \ep

\atop\scriptstyle r' \ge \ep} d^4 r' ~
\vevc{ \phifour (r') \phitwo (r) \phitwo (0)} \cr
&~~~= {\rm Sg} \int_{\scriptstyle |r' - r| \ge \ep

\atop\scriptstyle r'  \ge \ep} d^4 r'~
{1 \over 4}~\D(r')^2 \D(r-r')^2 \cr
&~~~= {\rm Sg} \int_{\scriptstyle |r' - r| \ge \ep

\atop\scriptstyle r'  \ge \ep} d^4 r'~
{1 \over 4 (4 \pi^2)^4}~{1 \over r'^4 |r-r'|^4} \cr
&~~~= {\ln r - \ln \ep \over 256 \pi^6 r^4} .\cr}}

\subsec{$k_{\x,m\x} (r;\ep)$}

This integral has dimension six, and it has

terms up to first order in $m^2$ ($v$ is counted
as first order in $m^2$)
by the dimensional analysis.  Hence, we can
expand the integrand up to first order in $m^2$.

We find, using \eopelist,
\eqn\ekxmxcal{\eqalign{&k_{\x,m\x} (r;\ep) \equiv {\rm Sg}
\int_{\scriptstyle |r' - r| \ge \ep

\atop\scriptstyle r' \ge \ep} d^4 r' ~
\vevc{\phifour (r') \left(\phitwo (r) \phifour (0) - C_{m\x,m} (r)
\phitwo (0) \right)} \cr
&~~~= {\rm Sg} \int_{\scriptstyle |r' - r| \ge \ep

\atop\scriptstyle r'  \ge \ep} d^4 r'~
{1 \over 6} \D(r')^3 \D(r'-r) \D(r) + {v \over 2} (k_{\x,mm} (r;\ep)
+ k_{m,m\x} (r;\ep)) \cr
&~~~= {\rm Sg} \int_{\scriptstyle |r' - r| \ge \ep

\atop\scriptstyle r'  \ge \ep} d^4 r'~
{1 \over \pi^{10}} \Bigg[ {1 \over 6144 r^2}~{1 \over r'^6
|r'-r|^2} \cr
&~~~~~+ m^2 \Bigg( \left( - {1 \over 24576} + {\ln r \over
12288}
\right) {1 \over r'^6 |r'-r|^2} + {1 \over r^2} \Big(
- {1 \over 24576} {1 \over r'^6} \cr
&~~~~~~~~ - {1 \over 8192} {1 \over r'^4 |r'-r|^2}
+ {1 \over 4096} {\ln r' \over r'^4 |r'-r|^2} +
{1 \over 12288} {\ln |r'-r| \over r'^6} \Big) \Bigg) \cr
&~~~~~+ \pi^2 v \Bigg( {1 \over 2048} {1 \over r'^4 |r'-r|^4}
+ {1 \over 1536} {1 \over r'^6 |r'-r|^2}
+ {1 \over r^2} \left( {1 \over 1536} {1 \over r'^6}
+ {1 \over 512} {1 \over r'^4 |r'-r|^2} \right) \Bigg) \Bigg] \cr
&~~~= {1 \over \pi^8} \Bigg[ {1 \over 6144 \ep^2 r^4}

+ {\ln \ep \over r^4} \left({5 m^2 \over 24576} - {3 \pi^2 v \over
512} \right) - {m^2 \ln^2 \ep \over 4096 r^4} - {1 \over 12288
r^6} \cr
&~~~~~+ {1 \over r^4} \left({7 m^2 \over 98304} +
{5 \pi^2 v \over 3072} + {3 \pi^2 v \over 512}~\ln r

+ {m^2 \over 4096}~\ln^2 r \right) \Bigg] . \cr}}

\subsec{$k_{\x,\x\x} (r;\ep)$}

The integral has dimension eight, and we must expand the
integrand up to second order in $m^2$.  We find, using
\eopelist,
\eqn\ekxxxcal{\eqalign{&k_{\x,\x\x} (r;\ep) \equiv

{\rm Sg} \int_{\scriptstyle |r' - r| \ge \ep

\atop\scriptstyle r' \ge \ep} d^4 r' \cr
&~~~\vevc{\phifour (r') \left(\phifour (r) \phifour (0)
- C_{\x\x,\x} (r) \phifour (0) - C_{\x\x,m} (r) \phitwo (0)

\right)} \cr
&~~= j_{\x,\x\x} (r;\ep) + v k_{\x,m\x} (r;\ep)
+ {v \over 2} k_{m,\x\x} (r;\ep)

- {v^2 \over 4} k_{\x,mm} (r;\ep) ,\cr}}
where $k_{m,\x\x} (r;\ep)$ is defined by \ekmsdef\ and
calculated as \ekms, and the integral $j_{\x,\x\x}$ is
defined by
\eqn\ejxxxdef{\eqalign{& j_{\x,\x\x} (r;\ep) \equiv
{\rm Sg} \int_{\scriptstyle |r' - r| \ge \ep

\atop\scriptstyle r'  \ge \ep} d^4 r' \cr
&~~~\Big[ {1 \over 8}~\D(r')^2 \D(|r'-r|)^2 \D(r)^2
- C_{\x\x,\x} (r) {1 \over 24}~\D(r')^4 \Big] .\cr}}

By expanding the integrand up to second order in $m^2$,
we obtain
\eqn\ejxxxcal{\eqalign{&j_{\x,\x\x} (r;\ep) =

{\rm Sg} \int_{\scriptstyle |r' - r| \ge \ep

\atop\scriptstyle r'  \ge \ep} d^4 r'~
{1 \over \pi^{12}}
\Bigg\lbrace {1 \over 32768 r^4} \left( - {1 \over r'^8} + {1
\over
r'^4 |r'-r|^4} \right) \cr
&~~~+ m^2 \Bigg[ {1 \over r^2} \left( - {1 \over 65536}
+ {\ln r \over 32768} \right) {1 \over r'^4 |r'-r|^4} +
{1 \over r^4} \Big( {1 \over 32768} {1 \over r'^6} \cr
&~~~~~- {1 \over 16384} {\ln r' \over r'^6} -
{1 \over 32768} {1 \over r'^4 |r'-r|^2} + {1 \over 16384}
{\ln |r'-r| \over r'^4 |r'-r|^2} \Big) \Bigg] \cr
&~~~+ \pi^2 v \Bigg[ {1 \over 4096 r^2} {1 \over r'^4 |r'-r|^4}
+ {1 \over r^4} \left( - {1 \over 2048} {1 \over r'^6}

+ {1 \over 2048} {1 \over r'^4 |r'-r|^2} \right) \Bigg] \cr
&~~~+ m^4 \Bigg[ \left( - {3 \over 1048576} - {\ln r \over
262144} + {\ln^2 r \over 131072} \right) {1 \over r'^4

|r'-r|^4} \cr
&~~~~~+ {1 \over r^2} \left( \left({1 \over 65536} -

{\ln r \over 32768} \right) {1 \over r'^4 |r'-r|^2} +
\left( - {1 \over 32768} + {\ln r \over 16384} \right)
{\ln |r'-r| \over r'^4 |r'-r|^2} \right) \cr
&~~~~~+ {1 \over r^4} \Big\lbrace - {1 \over 131072}
\left( {1 \over r'^4} - {1 \over r'^2 |r'-r|^2} \right)
+ {1 \over 131072} {\ln r'/|r'-r| \over r'^4} \cr
&~~~~~~~~~~+ {1 \over 32768} \ln r' \left({1 \over r'^4} - {1
\over
r'^2 |r'-r|^2} \right) \cr

&~~~~~~~~~~- {1 \over 65536} {\ln^2 r' - \ln^2 |r'-r|
\over r'^4} - {1 \over 32768} \left( {\ln^2 r' \over
r'^4} - {\ln r' \cdot \ln |r'-r| \over r'^2 |r'-r|^2} \right) \Big\rbrace
\Bigg] \cr
&~~~+ \pi^2 m^2 v \Bigg[ \left( - {1 \over 32768} + {\ln r
\over 8192} \right) { 1 \over r'^4 |r'-r|^4} \cr
&~~~~~+ {1 \over r^2} \left( {1 \over 2048} (-1 + \ln r)
{1 \over r'^4 |r'-r|^2} + {1 \over 2048} {\ln |r'-r|
\over r'^4 |r'-r|^2} \right) \cr
&~~~~~{1 \over r^4} \left(

\left( {1 \over 4096} - {\ln r' \over 2048} \right)

\left( {1 \over r'^4} - {1 \over r'^2 |r'-r|^2} \right)
- {1 \over 4096} {\ln r'/|r'-r| \over r'^4} \right) \Bigg] \cr
&~~~+ \pi^4 v^2 \Bigg[ {1 \over 2048} {1 \over r'^4 |r'-r|^4}
+ {1 \over 256 r^2} {1 \over r'^4 |r'-r|^2}
- {1 \over 512 r^4} \left( {1 \over r'^4}
- {1 \over r'^2 |r'-r|^2} \right) \Bigg] \Bigg\rbrace . \cr}}

\filbreak

Doing the integrals, we finally obtain, from \ekxxxcal,
\eqn\ekxxxresult{\eqalign{&k_{\x,\x\x} (r;\ep) = {1 \over
\pi^{10}} \Bigg\lbrace -{1 \over 65536 \ep^4 r^4}
- {1 \over \ep^2 r^4} \left( {\pi^2 v \over 3072} + {m^2 \ln \ep
\over 16384} \right) \cr
&~~~+ \ln \ep \Bigg[
- {1 \over 8192 r^8} + {1 \over r^6} \left(
{m^2 \over 8192} (1 - 2 \ln r) - {11 \pi^2 v \over
6144} \right) \cr
&~~~~~+ {1 \over r^4} \Big( - {m^4 \over 262144} +
{35 \pi^2 m^2 v \over 49152} - {9 \pi^4 v^2 \over
1024} \cr
&~~~~~~~+ m^2 \ln r \left( {m^2 \over 32768} - {11 \pi^2 v
\over 4096} \right) - {3 \over 16384}~m^4 \ln^2 r \Big) \Bigg]
\cr
&~~~+ {\ln^2 \ep \over r^4} \left( - {5 m^4 \over 131072}
+ {\pi^2 m^2 v \over 2048} \right)

+ {1 \over 32768}~m^4 \ln^3 \ep \cr
&~~~+ {\ln r \over 8192 r^8} + {1 \over r^6}
\Bigg[ - {m^2 \over 32768} + {\pi^2 v \over
2048} \cr
&~~~~~+ \ln r \left( - {m^2 \over 16384} +
{11 \pi^2 v \over 6144} \right) + {1 \over 4096}~m^2 \ln^2 r
\Bigg] \cr
&~~~+ {1 \over r^4} \Bigg[
{11 m^4 \over 262144} - {13 \pi^2 m^2 v \over 16384}
+ {17 \pi^4 v^2 \over 2048} \cr
&~~~~~+ \ln r \left( - {27 m^4 \over 262144} + {13 \pi^2 m^2 v
\over
12288} + {9 \pi^4 v^2 \over 1024} \right) \cr
&~~~~~+ \ln^2 r \left( - {3 m^4 \over 131072} + {9 \pi^2
m^2 v \over 4096} \right) + {5 \over 32768}~m^4 \ln^3 r \Bigg]
\Bigg\rbrace .\cr}}

\appendix{F}{The integrals $k_{\x,mm}$,
$k_{\x,m\x}$, and $k_{\x,\x\x}$ in terms of the beta
functions and finite counterterms}

{}From \eCmCx\ and \eH\ we can compute the first order
corrections to $C_m (r)$ and $C_\x (r)$ in terms of
the beta functions and finite counterterms.  Then, using
the subtractions \eAresult\ and the variational formulas
\edcmm, \edcmx, and \edcxx, we can obtain the
integrals $k_{\x,mm}$,
$k_{\x,m\x}$, and $k_{\x,\x\x}$ in terms of the beta
functions and finite counterterms.  We note that the
divergences
of the integrals are unambiguously determined, since
we know the divergent part of the subtractions given by
\eAresult.

We first find
\eqn\ekxmmexp{k_{\x,mm} (r;\ep) = {1 \over 256 \pi^6 r^4}
\left(
- \ln \ep + \ln r \right) ,}
which is completely determined.  We then find
\eqn\ekxmxexp{\eqalign{
& k_{\x,m\x} (r;\ep) = {1 \over 6144 \pi^8 \ep^2 r^4} +

{1 \over \pi^4 r^6} \left( \pi^2 \partial_\x c_{m\x,\one} (0,0)
- {1 \over 4} c_{\x\x,\R} (0) \right) \cr
&~+ {1 \over \pi^8 r^4} \Bigg(
\left( {5 m^2 \over 24576} - {3 \pi^2 v \over 512} \right) \ln \ep
- {m^2 \ln^2 \ep \over 4096} + {5 \pi^2 v \over 3072} \cr
&~~ + m^2 \left( - {1 \over 49152} + {c \over 32} + {\pi^4 \over
32}~
(2 c_{\x\x,\R} (0) + \partial_{m^2} c_{\x\x,m} (0,0)) \right)

+ {3 \pi^2 v \over 512}~\ln r + {m^2 \over 4096}~\ln^2 r \Bigg)
\cr}}
\tolerance=10000
\eqn\ekxxxexp{\eqalign{
& k_{\x,\x\x} (r;\ep) = {1 \over \pi^{10}} \Bigg\lbrace

- {1 \over 65536 \ep^4 r^4} - {1 \over \ep^2 r^4}
\left( {\pi^2 v \over 3072} + {m^2 \ln \ep \over 16384} \right) \cr
&~ + \ln \ep  \Bigg(
- {1 \over 8192 r^8 } + {1 \over r^6}
\left( {m^2 \over 8192} - {11 \pi^2 v \over 6144}

- {m^2 \ln r \over 4096} \right) \cr
&~~ + {1 \over r^4} \left( - {m^4 \over 262144}
+ {35 \pi^2 m^2 v \over 49152} - {9 \pi^4 v^2 \over 1024}
+ m^2 \ln r \left( {5 m^2 \over 32768} - {11 \pi^2 v \over 4096}
\right)
- {3 \over 16384}~m^4 \ln^2 r \right) \Bigg) \cr
&~ + {m^2 \ln^2 \ep \over r^4} \left( - {5 m^2 \over 131072}
+ {\pi^2 v \over 2048} \right) + {1 \over 32768 r^4}~m^4 \ln^3
\ep \cr
&~ + {1 \over r^8} \left( - {1 \over 32768} + 2 \pi^8 \partial_\x

c_{\x\x,\one} (0,0) + {\ln r \over 8192} \right) \cr
&~ + {1 \over r^6} \Bigg[ m^2 \left(
- {c \over 192} + {1 \over 16}~\pi^6 \partial_\x c_{m\x,\one}
(0,0)
- {1 \over 32}~\pi^6 \partial_\x c_{\x\x,m} (0,0) + \pi^8
\partial_\x
\partial_{m^2} c_{\x\x,\one} (0,0) \right) \cr

&~~ + \pi^2 v \left( - {11 \over 12288} + {1 \over 2}~\partial_\x
c_{\x\x,m} (0,0) - {1 \over 4}~c_{\x\x,R} (0) \right) \cr
&~~ + \ln r \left( m^2 \left(
- {1 \over 4096} - {1 \over 8}~\pi^6 \partial_\x c_{m\x,\one}
(0,0)
+ {1 \over 16}~c_{\x\x,m} (0,0) \right) + {11 \pi^2 v \over 6144}
\right)
+ {1 \over 4096}~m^2 \ln^2 r \Bigg] \cr
&~ + {1 \over r^4} \Bigg[

{\pi^8 v^2 \over 16}~b_{\x,2} + \pi^6 m^2 v \left( {\pi^2 \over
4}~b_{m,3}
+ {1 \over 16}~c_{\x\x,\R} (0)

+ {1 \over 4}~\partial_{m^2} c_{\x\x,m} (0,0) \right) \cr
&~~ + m^4 \Big( {5 c \over 512} + {\pi^4 c \over 4}~b_{\x,2} +
{\pi^6 \over 16}~\partial_{m^2} c_{m\x,\one} (0,0) + {\pi^8
\over 2}~
b_{\one,3} + {5 \pi^6 \over16}~ \partial_{m^2}^2 c_{\x\x,\one}
(0,0) \cr
&~~ - {\pi^6 \over 32}~\partial_\x \partial_{m^2} c_{\x\x,m}
(0,0) \Big)

+ \ln r \Big\lbrace m^4 \left( {1 \over 131072} - {c \over 256}

+ {\pi^6 \over 32}~b_{m,3} + {7 \pi^4 \over
256}~\partial_{m^2} c_{\x\x,m}
(0,0) \right) \cr
&~~ + \pi^2 m^2 v \left( - {25 \over 24576} + {\pi^4 \over
64}~b_{\x,2}
\right) + {9 \pi^4 v^2 \over 1024} \Big\rbrace \cr
&~ + \ln^2 r \Big\lbrace

m^4 \left( - {5 \over 32768} + {\pi^4 \over 1024}~b_{\x,2}
\right)
+ {9 \pi^2 m^2 v \over 4096} \Big\rbrace

+ {5 m^4 \over 32768}~\ln^3 r
\Bigg] \Bigg\rbrace . \cr}}

\listrefs
\parindent=20pt
\bye